# Ultrathin Nanostructured Metals for Highly Transmissive Plasmonic Subtractive Color Filters

Beibei Zeng, Yongkang Gao, Filbert J. Bartoli[*]

*Electrical and Computer Engineering Department, Lehigh University, Bethlehem, PA 18015*

*\* fjb205@lehigh.edu*

**Abstract** Plasmonic color filters employing a single optically-thick nanostructured metal layer have recently generated considerable interest as an alternative to colorant-based color filtering technologies, due to their reliability, ease of fabrication, high color tunability. However, their relatively low transmission efficiency (~30%) is an important challenge that needs to be addressed for practical applications. The present work reports, for the first time, a novel plasmonic subtractive color filtering scheme that exploits the counter-intuitive phenomenon of extraordinary low transmission (ELT) through an ultrathin nanostructured metal film. This approach relies on a fundamentally different color filtering mechanism than that of exsiting plasmonic additive color filters, and achieves unusually high transmission efficiencies of 60~70% for simple architectures. Furthermore, owing to short-range interactions of surface plasmon polaritons at ELT resonances, our design offers high spatial resolution color filtering with compact pixel size close to the optical diffraction limit ($\sim\lambda/2$), creating solid applications ranging from imaging sensors to color displays.

Color filters are vital components for digital photography, projectors, displays, image sensors and other optical measurement instrumentation. Traditional on-chip color filters that employ organic dyes or chemical pigments are vulnerable to processing chemicals, and suffer from performance degradation under long-duration ultraviolet irradiation or at high temperatures. Furthermore, highly-accurate lithographic alignment techniques are required to pattern each type of pixel over a large area, increasing fabrication difficulty and cost [1-3]. One promising approach to overcome these challenges is the use of plasmonic color filters [1-10]. For instance, a single optically-thick metal layer perforated with periodic subwavelength hole arrays can exhibit the well-known extraordinary optical transmission (EOT) phenomenon [11-13], which has been extensively studied for color filtering over the past decade. These plasmonic additive color filters (ACFs) block the whole visible spectrum except for selective transmission bands that are associated with the excitation of surface plasmon polaritons (SPPs) [1,2,5-13]. These EOT transmission bands can be spectrally tuned throughout the entire visible spectrum by simply adjusting geometric parameters, such as the periodicity, shape and size of the holes, leading to the wide color tunability. Single-layer metal nanostructures also have advantages over colorant-based materials due to their ease of fabrication and device integration, and greater reliability under high temperature, humidity and long-term radiation exposure [1,6-8]. In spite of these significant advantages, the low transmission efficiency of hole-array plasmonic ACFs (~30% at visible wavelengths) remains as a bottleneck that limits their commercial applications [7]. Recently, peak transmission efficiencies of 40~50% were achieved in the state-of-art hole-array plasmonic ACFs by matching the refractive indices of media on both sides of the metal film [1], which, however, is still far below that for commercial image sensors (~80%, FUJIFILM Electronic materials U.S.A., Inc.). Plasmonic ACFs formed by metal-insulator-metal (MIM) or metal-dielectric (MD) waveguide nanoresonators have recently achieved high transmission efficiencies of 50~80% [2,9,10]. Nevertheless, these complex multilayer designs are not suitable for cost-effective nanofabrication and device integration. In this context, it is highly desirable to design novel plasmonic color filters with both high transmission efficiency and simple architectures.

The present work explores the counter-intuitive extraordinary low transmission (ELT) phenomenon [14-20], and report a novel approach to achieving plasmonic subtractive color filters (SCFs) with unusually high transmission of 60~70% in a single ultrathin (30*nm*) metal film patterned with one dimensional (1D) nanogratings. The transmission minima of plasmonic SCFs, corresponding to the ELT resonances, can be tuned to specific frequencies across the entire visible region by simply varying the period of nanogratings. In this subtractive color filtering scheme, specific colors (i.e. cyan, magenta, and yellow, CMY) are generated by removing their complementary components (i.e. red, green, and blue, RGB) from the visible spectrum. Due to their broad passbands with twice the photon throughput of narrowband RGB ACFs, CMY SCFs have the major advantages of higher quantum efficiency and stronger spectral response, and have been successfully used in image sensors for years [21-23]. Unfortunately, highly efficient plasmonic SCFs were not possible using exsiting techniques and have not been previously realized. The present work exploits recent advances in thin-film plasmonic nanostructures, and achieves for the first time, plasmonic SCFs with high transmission efficiency close to that for commercial image sensors. Moreover, owing to short-range interactions of SPPs between nearest-neighbor nanostructures in the ELT resonance, the plasmonic SCFs can efficiently filter colors with only a few (even two) nanoslits, yielding ultracompact pixel sizes close to the optical diffraction limit (~$\lambda$/2, i.e. 200~350*nm*) that determines the highest possible optical resolution [3,24]. Therefore, plasmonic SCFs are capable of generating even smaller pixel sizes than the smallest pixels achieved today in commercial image sensors (1.12×1.12$\mu m^2$, Sony Corp.). Finally, the color filtering functions of plasmonic SCFs are polarization-dependent, which can either filter transverse-magnetic (TM) polarized illumination, or function as highly transparent windows under transverse-electric (TE) polarization. These unique properties make the ultrathin plasmonic SCFs highly attractive for high-definition transparent displays [25,26].

**Results**

**Ultrathin plasmonic subtractive color filters basd on Extraordinary Low Transmission**

Figure 1 (a) is a photograph of a 30*nm*-thick Ag film deposited on a standard microscope glass slide. The background pattern can be clearly seen through the semi-transparent Ag film. The Ag film thickness was determined to be 29.8*nm*, and its measured optical constants are noticeably different from those of an optically thick (350*nm*) Ag film (Supplementary Figure S1). A schematic diagram of the proposed ultrathin plasmonic SCFs is shown in Figure 1 (b), where 1D nanogratings with different periods are patterned on the 30*nm*-thick Ag film. For normally incident light polarized along the *x*-direction (TM polarization), the two single-interface (air/metal and metal/glass interfaces) SPPs couple to form strongly damped short-range SPPs (SRSPPs) in this asymmetric geometry [14-20]. The absorption and reflection is enhanced at the SRSPP resonance wavelength [20], leading to a transmission dip, which is the complement of the well-known EOT phenomena that exhibits enhanced transmission peak at the resonance wavelength [11-13]. Consequently, arbitrary colors may be subtracted from broadband white light by simply varying the period of nanopatterns on the ultrathin metal film to tune the SRSPP resonance wavelength. Details of the physical mechanisms will be discussed later. The key features of such designs, which contain only a single nanopatterned ultrathin metal layer, are their simple design rules, ease of fabrication, and scalable throughput by means of large-area nanofabrication methods, such as nanoimprint lithography or optical interference lithography [27-29]. For a proof-of-principle experiment, nanogratings with different periods were fabricated using focus ion beam milling. The right column in Figure 1 (c) shows scanning electron microscopy (SEM) images of the fabricated nanogratings with three different periods (230*nm*, 270*nm* and 350*nm*). The duty cycle of the nanogratings is set as 0.5. Figure 1 (c) presents measured transmission spectra of the cyan (P=350*nm*), magenta (P=270*nm*) and yellow (P=230*nm*) plasmonic SCFs (Supplementary Figure S2), with transmission dips that are positioned in red, green, and blue spectral regions, respectively. Note that the observed absolute transmission peaks, 60~70% in the visible region, represent an unusually high transmission efficiency for such structures [1,2,6-13]. The full-width at half maximum (FWHM) of the stopbands is about 100 *nm* for yellow and cyan SCFs, and 160*nm* for magenta SCFs, which are comparable to the passband bandwidth reported for state-of-art plasmonic ACFs [1,2].

In order to achieve a full palette of subtractive colors that spans the entire visible region, we have continuously varied the period of nanogratings from 220*nm* to 360*nm*, in 10*nm* increment. All the fabricated nanogratings have the same dimensions of $10\times10\mu m^2$. Figure 2 (a) shows the corresponding optical microscope images (from yellow to cyan) of fifteen square-shaped plasmonic SCFs illuminated by TM-polarized white light (Supplementary Figure S2). At the same time, these nanostructures strongly transmit TE-polarized light (Supplementary Figure S3), which distinctly contrasts with that of previous plasmonic ACFs or wire-grid polarizers [2,30]. The polarization-dependent of color filtering effects in plasmonic SCFs arise from the polarization-dependent excitation of SRSPPs in 1D nanogratings. This unique feature indicates that the proposed plasmonic SCFs can function as SCFs or highly transparent windows under different polarization states, which has potential applications in transparent displays [25,26]. Figure 2 (b) presents transmission spectra of the plasmonic SCFs shown in Figure 2 (a), with the spectra exhibiting dips that are tuned across the visible spectrum by varying the period from 220*nm* to 360*nm*. Finite-difference time-domain (FDTD) numerical simulations (Fig. 2 (b), i) agree reasonably well with the experimental results (Fig. 2 (b), ii). The trend lines (black dashed lines) approximate the variation of transmission dips from 470*nm* to 620*nm* as the periods vary from 220*nm* to 360*nm*. The transmission dips, which correspond to SRSPP resonance wavelengths, are further illustrated in Figure 2 (c), showing a nearly linear relation between the resonance wavelengths and periods of nanogratings. It is highly advantageous that arbitrary subtractive colors can be obtained by simply varying the period of nanogratings in the proposed plasmonic SCFs, which could extend the operating range of conventional colorant color filters that do not scale well to more than three spectral bands, making them especially attractive for multispectral imaging applications [5].

To understand the physical mechanisms more clearly, we model the optical properties of plasmonic SCFs by FDTD numerical simulation. 2D maps of the calculated transmission and absorption for 30*nm*-thick Ag nanogratings are shown in Figure 3 (a) and (b), respectively, as a function of the incident wavelength and grating period. The duty cycle of nanogratings is also set as 0.5. Figure 3 (a) clearly shows a broad transmission minimum associated with SRSPPs, whose resonance wavelength varies continuously from 400*nm* to 650*nm* in wavelength as the period increases from 100*nm* to 400*nm*. A narrow transmission peak attributed to Rayleigh-Wood anomaly at the Ag/glass interface (indicated by the black solid line) ranges from 300*nm* to almost 600*nm* in wavelength as the grating period increases from 200*nm* to 400*nm* [31]. The black dashed line in Figure 3 (a) depicts the character of Rayleigh-Wood anomaly at the air/Ag interface, which matches well with the transmission peak in the ultraviolet region (300~400*nm*). Both enhanced optical reflection introduced by re-radiation of the SRSPP mode and absorption in the nanopatterned ultrathin Ag film play an important role in suppressing transmission at the resonance wavelength. The dispersion characteristics in Figure 3 (a) and (b) show that the transmission minimum is primarily due to enhanced reflection for larger periods and to absorption in the Ag nanograting for periods less than 250*nm*. The white solid and dash-dotted curves in Figure 3 (a) and (b) represent analytical dispersion relations for the lowest and higher orders SRSPP modes, respectively, in good agreement with the optical transmission dips in Figure 3 (a) and absorption peaks in Figure 3 (b) obtained by FDTD numerical simulation.

In Figure 3 (c), the cyan and black solid curves are the measured transmission spectra under TM-polarized illumination through 30*nm*-thick Ag films with and without nanogratings of period P=340*nm*, respectively. An obvious transmission dip is centered at a wavelength of 610*nm*, which agrees well with the numerical results represented by cyan dashed curves. The intriguing ELT phenomenon is clearly observed: the optical transmission through the ultrathin Ag film patterned with nanogratings (cyan solid or dashed curves) is lower than that through the solid Ag film without nanogratings (black solid or dashed curves) over a broad spectral range (gray region), even though 50% Ag is removed in the nanogratings compared to the unpatterned Ag film. It should be noted that the minimum transmission is zero in the simulation, indicating that the incident light could be completely blocked by the ultrathin nanogratings at the resonance wavelength. The difference between the experimental and numerical results can be attributed to the nonparallel incident light employed in the optical measurement, nanofabrication errors and surface roughness, which are not considered completely in numerical simulations. To further characterize SRSPP modes at the resonance wavelength, we calculate the $E_z$ vector distribution at the air/Ag and glass/Ag interfaces for nanogratings (P=340*nm*) at the wavelength of 610*nm* (indicated by the red-cross in Figure 3 (a)), and plot the results in Figure 3 (d). The antisymmetric $E_z$ field patterns correspond to a symmetric surface charge distribution (Supplementary Figure S4), demonstrating that the resonant electromagnetic modes are SRSPPs [32]. Additional simulations reveal that the electromagnetic modes in a relatively broad spectral region close to the transmission dip have similar field distributions. Consequently, it is the SRSPP modes excited in the ultrathin Ag nanogratings that cause the ELT phenomenon.

**High-resolution plasmonic subtractive color filtering and applications**

Next, we investigate the functional relationship between color filtering and the feature size of plasmonic SCFs, and attempt to determine the smallest possible pixel size for imaging applications. Figure 4 (a) shows cyan and magenta plasmonic SCF arrays consisting of 2, 4, 6, 8 and 10 nanoslits with the same length of 15 $\mu m$. The nanoslit periods for the cyan and magenta SCFs are 350$nm$ and 270$nm$, respectively. The duty cycle is set as 0.5, as before. Surprisingly, SCF arrays with only two nanoslits can still exhibit distinct cyan (ii) or magenta (iii) colors, indicating that the ultra-compact architectures are feasible. Here, the nanoscale dimensions (525$nm$ and 405$nm$ for the cyan and magenta filters with two nanoslits, respectively) are close to the diffraction limit of visible light ($\lambda/2$, 200~350$nm$) [3,24]. Moreover, as shown in Figure 4 (b), the color filtering effect of cyan and magenta plasmonic SCFs with a few nanoslits can be potentially maintained even though the length of nanoslits is decreased to 300$nm$ (Supplementary Figure S5). This interesting behavior can be understood in terms of the strong confinement properties of SRSPP modes [14-20,32]. Most of the electric field of SRSPPs is concentrated in the metal film, resulting in strong Ohmic losses and short propagation distances that are less than 1$\mu m$ at visible wavelengths [16]. The short propagation distance of SRSPPs makes interactions between nanostructures that are not nearest neighbors much weaker than those for EOT phenomenon, where SPPs excited at each nanoslit (or nanohole) strongly interfered with numerous nearby nanoslits (or nanoholes) [11-13]. Therefore, the multiple repeat units that are commonly employed in plasmonic ACFs based on EOT theory are not required in plasmonic SCFs due to short-range interactions of SRSPPs. Note that a unique feature of the plasmonic SCFs is their ability to perform color filtering on the nanometer scale, with much simpler and six times thinner structures than that of previous multilayered designs [2].

In addition, Figure 4 (c) shows a 2×2 array of plasmonic SCFs designed to investigate the effect of spatial crosstalk on transmitted colors between adjacent structures. A color filter mosaic was fabricated consisting of four different square-shaped ($10 \times 10 \mu m^2$) plasmonic SCFs with zero separation, in which four SCFs are composed of nanogratings with different periods ($P_1=220nm$, $P_2=260nm$, $P_3=290nm$, and $P_4=350nm$), as shown in the top panel of Figure 4 (c). The optical microscope image of the color filter mosaic under TM-polarized white light is shown in the bottom panel of Figure 4 (c). Four distinct subtractive colors can be clearly resolved even at the center corner or boundaries of adjacent filters, indicating that the proposed plasmonic SCFs can be applied to high-resolution color filter arrays (CFAs) widely used in imaging sensors or color displays [21-23,33]. The image blurring at boundaries arises from effects of light diffraction and the limited optical resolution of the microscope.

Spectral imaging combines two methodologies, spectroscopy and imaging, for applications ranging from biological studies to remote sensing. However, this technique typically employs bulky filters or interferometers (with multispectral bands) combined with scanning to acquire a complete spectrum at each pixel [23], since conventional miniature CFAs are limited to three spectral bands (i.e. RGB or CMY) [2]. In order to enable direct recording of spectral image data in a single exposure without scanning, the plasmonic photon sorters (with limited transmission efficiency 1.5~15% [5]) and ultra-compact plasmonic spectroscope (composed of complex MIM nano-resonators [2]) were proposed for achieving miniature CFAs with multispectral bands due to their wide color tunability. We also design a plasmonic subtractive spectroscope simply formed by plasmonic SCFs arrays. Figure 5 (a) shows a SEM image of the fabricated device consisting of ultrathin nanogratings with periods gradually changing from 220*nm* to 360*nm* in increments of 1*nm* and a fixed linewidth of 110*nm*. When illuminated with TM-polarized white light, the structure becomes a rainbow stripe of continuous subtractive colors, as shown in Figure 5 (b). The miniature plasmonic subtractive spectroscope can split the entire visible spectrum into component colors within a distance of a few micrometers, which are orders of magnitude smaller than the conventional prism- or grating-based devices for multispectral imaging [23]. Moreover, the plasmonic subtractive spectroscope has a much higher transmission efficiency (60~70%), a simple scheme consisting of a single nanopatterned ultrathin metal film, and is five to ten times thinner than that of previous designs [2,5].

Next, we demonstrate the potential of plasmonic SCFs for transparent displaying [25,26]. Figure 5 (c,i) shows an optical microscope image of a magenta character 'L' in a cyan background, formed when nanopatterns are illuminated with TM-polarized white light. The letter 'L' is constructed by nanogratings with a period of P=270*nm*, and the background by nanogratings with a period of P=350*nm* (Supplementary Figure S6). Two distinct colors can be clearly preserved even at the sharp corners and boundaries of two different patterns, indicating the high-resolution color filtering capability. On the other hand, for TE polarization, the same structure becomes a transparent window with high transmission, through which we can clearly observe a background object with its detailed features, as shown in Figure 5 (c,ii). This is quite different from that of the plasmonic nanoresonators ACFs, for which the TE-polarized incident light is totally blocked [2]. Therefore, the ultrathin plasmonic SCFs can function as color filters as an alternative to conventional colorant CFs or plasmonic ACFs, or act as a highly transparent window under illumination with a different polarization, offering a new approach for high-definition transparent displays through actively controlling the polarization of incident light at each color pixel.

**Discussion**

In solid-state CMOS and CCD image sensor technologies, the main method of implementing color imaging is to insert color filters in the optical path. The conventional colorant color filters suffer from several drawbacks as mentioned before. An alternative method is to use plasmonic ACFs formed by periodic nanohole arrays patterned on a single optically-thick metal film. This technique produces the required color filters in a single metal layer, which reduce both cost and color cross-talk. And such plasmonic ACFs can be easily integrated in the lower metal layers that are close to photodiodes [34,35]. However, the low transmission efficiency of hole-array plasmonic ACFs restricts their commercial applications for image sensors. On the other hand, image sensors have employed SCFs instead of ACFs to improve the quantum efficiency and spectral response, because SCFs have superior light transmission characteristics [21,22]. Therefore, the proposed plasmonic SCFs with high transmission, simple designs and ultra-compact pixel sizes could be a great alternative to colorant color filters and plasmonic ACFs.

In addition, picture quality and color richness are the most important considerations in color displays. Standard computer monitors and television displays are typically formed by reproduction of three primary RGB colors. Since the natural objects are more colorful than the current displays, there is an urgent need to produce wider color gamut in order to reproduce the original colors with high fidelity [33]. Therefore, displays with multi-colors (more than RGB) have already been developed to widen the reproducible color gamut [33]. The plasmonic SCFs could be employed, in combination with plasmonic ACFs [36], to establish sufficient coverage of a desired color gamut due to their wide color tunability. These plasmonic color filters could provide color pixels with dimensions of only several micrometers, or less, which are much smaller than the resolution limit of human eyes (~80$\mu m$ at a 35 mm distance) [37]. Furthermore, the longitudinal thickness of plasmonic color filters is 1-2 orders of magnitude smaller than that of colorant ones, which would be useful in the design of ultrathin display panels [2]. Finally, the unique polarization-dependent feature of the proposed plasmonic SCFs, working as either color filters under TM polarization or highly-transparent windows under TE polarization, is attractive for developing high-definition transparent displays.

It is also worth noting that fabrication challenges remain, since large grain sizes and surface roughness in ultrathin (30*nm*) Ag films could significantly affect the performance of plasmonic SCFs and lead to measurement errors and non-uniform colors. However, improved SCF structures can be realized by introducing an intermediate (1*nm*) Ge wetting layer before depositing Ag on the glass substrate [38], permitting ultrasmooth 30*nm*-thick Ag films with smaller grain sizes for improved color filtering performance.

In summary, by exploiting ELT theory, we have proposed and demonstrated plasmonic SCFs associated with fundamentally different color filtering mechanisms than previous state-of-art plasmonic ACFs. The simple design, with wide color tunability, ease of fabrication and device integration and also great reliability, combines advances of SCFs and plasmonic nanostructures to overcome the key challenges in current colorant and plasmonic color filters. Unusually high transmission efficiency of 60~70% has been achieved and can still be enhanced in the future. In addition, the proposed plasmonic SCFs were capable of generating even smaller pixel sizes than the smallest pixels achieved today in commercial image sensors. Also, their unique polarization-dependent features allow the same structures to function either as color filters or highly-transparent windows under different polarizations, opening an avenue towards high-definition transparent displays. While only one-dimensional nanograting structures have been demonstrated, this design principle can be extended to two dimensional structures to achieve polarization-independent subtractive color filtering. The design can also be easily applied to other spectrum regimes for different applications.

## Methods

**Analytical solution of SRSPP dispersions in the ultrathin Ag film patterned with subwavelength periodic structures.** For TM-polarized incident light, the two single-interface SP modes at the top and bottom interfaces of ultrathin (30*nm*) Ag nanogratings would interact with each other and lead to coupled SP modes, the long-range and short-range surface plasmon polariton (LRSPPs and SRSPPs) modes. The dispersions for LRSPPs and SRSPPs modes can be described by the following equation [17]:

$$\tanh(k_2 t)(\varepsilon_{d1}\varepsilon_{d2}k_2^2 + \varepsilon_m^2 k_1 k_3) + \varepsilon_m k_2 (\varepsilon_{d1} k_3 + \varepsilon_{d2} k_1) = 0 \qquad (1)$$

Here $k_1^2 = k_{spp}^2 - \varepsilon_{d1} k_0^2$, $k_2^2 = k_{spp}^2 - \varepsilon_m k_0^2$, $k_3^2 = k_{spp}^2 - \varepsilon_{d2} k_0^2$, $k_0 = \omega/c$ and $t$ is the thickness of the metal film. $\varepsilon_{d1}$ and $\varepsilon_{d2}$ are dielectric constants of air and glass, and $\varepsilon_m$ represent dielectric constants of ultrathin (30*nm*) Ag films. For the ultrathin Ag film with an asymmetric geometry ($\varepsilon_{d1} < \varepsilon_{d2}$), Eq. (1) yields strongly damped SRSPP modes with antisymmetric Ez field patterns at the air/Ag and glass/Ag interfaces. The momentum mismatch between SP modes and free space light can be bridged by the reciprocal vectors of periodic nanostructures $k_G = mG$ ($G = 2\pi/P$, $P$ is the period, $m$ is an integer):

$$k_{spp} = k_0 \sin\theta + mG \qquad (2)$$

where $\theta$ is the incident angle. For the normal incidence ($\theta = 0^0$), the dispersions of SRSPP modes are obtained by substituting Eq. (2) into Eq. (1), as white solid and dash-dotted curves in Figure 3 (a) and (b) shows, representing analytical dispersion relations for the lowest and higher orders SRSPP modes, respectively.

**Device fabrication and optical measurement.** Ag films of 30*nm* thickness were deposited by e-beam evaporation (Indel system) onto standard microscope slides (Fisherbrand), with a deposition rate of 0.1nms$^{-1}$. Prior to the evaporation, the glass slides were cleaned thoroughly with acetone in an ultrasonic cleaner for 20min, followed by extensive DI water rinsing. Focused ion beam (FEI Dual-Beam system 235) milling (30kV, 30pA) was used to fabricate the nanogratings on the Ag films.

To investigate the optical properties of the fabricated structures, the observation setup and the measurement of transmission spectra are based on an Olympus IX81 inverted microscope (Supplementary Figure S2). A 100W halogen lamp was used as the white light source. The transmission light was collected by a 40× microscope objective with a numerical aperture of 0.6. The microscope field diaphragm and aperture stop were both closed in order to have approximately collimated incident light. The collected light was coupled into a multimode fiber bundle interfaced with a compact spectrometer (Ocean Optics USB 4000). And all optical images were taken with a digital camera (Cannon EOS Rebel T3i).

**Numerical simulations.** Simulations of the transmission or absorption spectra were carried out using the three-dimensional finite-difference time-domain (FDTD) method of the Lumerical commercial software package (Lumerical Solutions Inc.). The wavelength-dispersion permittivity of the ultrathin (30*nm*-thick) Ag film was measured using spectroscopic ellipsometer (J. A. Woolam), and incorporated into FDTD simulations.

# References

1. Yokogawa, S., Burgos, S. P. & Atwater, H. A. Plasmonic color filters for CMOS image sensor applications. *Nano Lett.* **12**, 4349 (2012).
2. Xu, T., Wu, Y-K., Luo, X. & Guo, L. J. Plasmonic nanoresonators for high resolution colour filtering and spectral imaging. *Nat. Commun.* **1**, 1 (2010).
3. Wu, Y-K., Hollowell, A. E., Zhang, C. & Guo, L. J. Angle-Insensitive Structural Colours based on Metallic Nanocavities and Coloured Pixels beyond the Diffraction Limit. *Sci. Rep.* **3**, 1194 (2013).
4. Diest, K., Dionne, J. A., Spain, M. & Atwater, H. A. Tunable color filters based on metal–insulator–metal resonators. *Nano Lett.* **9**, 2579 (2009).
5. Laux, E., Genet, C., Skauli, T. & Ebbesen, T. W. Plasmonic photon sorters for spectral and polarimetric imaging. *Nat. Photonics* **2**, 161 (2008).
6. Lee, H. S., Yoon, Y. T., Lee, S. S., Kim, S. H. & Lee, K. D. Color filter based on a subwavelength patterned metal grating. *Opt. Express* **15**, 15457 (2007).
7. Chen, Q. & Cumming, D. R. High transmission and low color cross-talk plasmonic color filters using triangular-lattice hole arrays in aluminum films. *Opt. Express* **18**, 14056 (2010).
8. Inoue, D. *et al.* Polarization independent visible color filter comprising an aluminum film with surface-plasmon enhanced transmission through a subwavelength array of holes. *Appl. Phys. Lett.* **98**, 093113 (2011).
9. Kaplan, A. F., Xu, T. & Guo, L. J. High efficiency resonance-based spectrum filters with tunable transmission bandwidth fabricated using nanoimprint lithography. *Appl. Phys. Lett.* **99**, 143111 (2011).
10. Yoon, Y. T., Park, C. H. & Lee, S. S. Highly efficient color filter incorporating a thin metal-dielectric resonant structure. *Appl. Phys. Express* **5**, 022501 (2012).
11. Ebbesen, T. W., Lezec, H. J., Ghaemi, H. F., Thio, T. & Wolff, P. A. Extraordinary optical transmission through subwavelength hole arrays. *Nature* **391**, 667 (1998).
12. Barnes, W. L., Dereux, A. & Ebbesen, T. W. Surface plasmon subwavelength optics. *Nature* **424**, 824 (2003).
13. Genet, C. & Ebbesen, T. W. Light in tiny holes. *Nature* **445**, 39 (2007).

14. Spevak, I. S., Nikitin, A. Y., Bezuglyi, E. V., Levchenko, A. & Kats, A. V. Resonantly suppressed transmission and anomalously enhanced light absorption in periodically modulated ultrathin metal films. *Phys. Rev.* B **79**, 161406 (2009).

15. Rodrigo, S. G., Martin-Moreno, L., Nikitin, A. Y., Kats, A. V., Spevak, I. S. & Garcia-Vidal, F. J. Extraordinary optical transmission through hole arrays in optically thin metal films. *Optics Lett.* **34**, 4 (2009).

16. D'Aguanno, G., Mattiucci, N., Alu, A. & Bloemer, M. J. Quenched optical transmission in ultrathin subwavelength plasmonic gratings. *Phys. Rev.* B **83**, 035426 (2011).

17. Braun, J., Gompf, B., Kobiela, G. & Dressel, M. How holes can obscure the view: suppressed transmission through an ultrathin metal film by a subwavelength hole array. *Phys. Rev. Lett.* **103**, 203901 (2009).

18. Xiao, S. *et al.* Nearly zero transmission through periodically modulated ultrathin metal films. *Appl. Phys. Lett.* **97**, 071116 (2010).

19. Xiao, S. & Mortensen, N. A. Surface-plasmon-polariton-induced suppressed transmission through ultrathin metal disk arrays. *Opt. Lett.* **36**, 37 (2011).

20. Gan, Q. *et al.* Short Range Surface Plasmon Polaritons for Extraordinary Low Transmission Through Ultra-Thin Metal Films with Nanopatterns. *Plasmonics* **7**, 47 (2012).

21. Nabeyama, H., Nagahara, S., Shimizu, H., Noda, M. & Masuda, M. All solid state color camera with single-chip MOS imager. *IEEE Trans. Consumer Electron.* **CE-27**, 40 (1981).

22. Sencar, H. T. & Memon, N. *Digital Image Forensics: Advances and Challenges* (Springer, New York, 2012).

23. Garini, Y., Young, I. T. & McNamara, G. Spectral imaging: principles and applications. *Cytometry* A **69A**, 735 (2006).

24. Kumar, K., Duan, H., Hegde, R. S., Koh, S. C. W., Wei, J. N. & Yang, J. K. W. Printing colour at the optical diffraction limit. *Nat. Nanotech.* **7**, 557 (2012).

25. Seo, H.-S. Transparent display apparatus. U.S. Patent No. 8227797 (2012).

26. Azuma, R., Baillot, Y., Behringer, R., Feiner, S., Julier, S. & MacIntyre, B. Recent Advances in Augmented Reality. *IEEE Comput. Graph. Appli.* **21**, 34 (2001).

27. Chou, S. Y., Krauss, P. R. & Renstrom, P. J. Imprint lithography with 25-nanometer resolution. *Science* **272**, 85 (1996).

28. Luo, X. & Ishihara, T. Surface plasmon resonant interference nanolithography technique. *Appl. Phys. Lett.* **84**, 4780 (2004).

29. Zeng, B., Yang, X., Wang, C. & Luo, X. Plasmonic interference nanolithography with a double-layer planar silver lens structure. *Opt. Express* **17**, 16783 (2009).

30. Wang, J. *et al.* High-performance nanowire-grid polarizers. *Opt. Lett.* **30**, 195 (2005).

31. Wood, R. W. On a remarkable case of uneven distribution of light in a diffraction grating spectrum. *Philos. Mag.* **4**, 396 (1902).

32. Chen, Z., Hooper, I. R. & Sambles, J. R. Strongly coupled surface plasmons on thin shallow metallic gratings, *Phys. Rev.* B **77**, 161405 (2008).

33. Cheng, H.-C., Ben-David, I. & Wu, S.-T. Five-Primary-Color LCDs. *J. Display Technol.* **6**, 3 (2010).

34. Catrysse, P. B. & Wandell, B. A. Integrated color pixels in 0.18-µm complementary metal oxide semiconductor technology. *J. Opt. Soc. Am.* A **20**, 2293 (2003).

35. Chen, Q. *et al.* A CMOS image sensor integrated with plasmonic colour filters. *Plasmonics* **7**, 695 (2012).

36. Yoon, M.-S. Method for fabricating color filter using surface plasmon and method for fabricating liquid crystal display device. U.S. Patent No. 7989254 (2011).

37. Betancourt, D. & del Rio, C. Study of the human eye working principle: an impressive high angular resolution system with simple array detectors. *IEEE Sensor Array Workshop* **93** (2006).

38. Logeeswaran, V. J. *et al.* Ultrasmooth Silver Thin Films Deposited with a Germanium Nucleation Layer. *Nano Lett.* **9**, 178 (2009).

**Figure 1. Plasmonic subtractive color filters formed by patterning nanogratings on ultrathin Ag films.** (a) A photograph of a 30*nm*-thick semi-transparent Ag film deposited on a microscope glass slide, showing a Lehigh University logo in the background. (b) Schematic diagram of the proposed plasmonic SCFs. The white light is incident from the glass side, and the cyan, magenta, purple and yellow beams represent the transmitted colored light through nanogratings with different periods patterned on the 30*nm*-thick Ag film. (c) Measured transmission spectra for yellow, magenta and cyan plasmonic SCFs consisting of ultrathin (30*nm*) nanogratings with periods of 230*nm*, 270*nm* and 350*nm*, respectively. Right column shows SEM images of the fabricated nanogratings with three different periods (230*nm*, 270*nm* and 350*nm*). Scale bars are 1*μm*.

**Figure 2. Optical micrographs and spectral analyses of ultrathin plasmonic subtractive color filters with varying periods.** (a) The full palette of transmitted subtractive colors from yellow to cyan is revealed in above 10×10*μm²* squares under TM-polarized white light, as the period of nanogratings in each square changes from 220*nm* to 360*nm* in 10*nm* increment. Scale bar is 5*μm*. (b) Simulated (i) and experimental (ii) transmission spectra of nanogratings with periods from 220*nm* to 360*nm*. The trend lines (black dashed lines) approximate the movement of the transmission dips with varying the periods of nanogratings. (c) Correlation between transmission dips observed in the experimental (red square) and simulation (black triangle) data.

**Figure 3. Theoretical and experimental verification of extraordinary low transmission for plasmonic subtractive color filtering.** 2D maps of the calculated TM optical transmission (a) and absorption (b) spectra of 30*nm*-thick Ag nanogratings as a function of the incident wavelength and grating period. The black solid and dashed lines in (a) refer to Rayleigh-Wood anomaly at glass/Ag and air/Ag interfaces, respectively. The white solid curve in (a) or (b), and dash-dotted curve in (b) correspond to the analytical dispersion relations for the lowest and higher order SRSPP modes, respectively. (c) Measured (i) and simulated (ii) TM transmission spectra through 30*nm*-thick Ag films with or without nanogratings of period P=340*nm*. The ELT phenomenon occurs in the gray spectral region. (d) Instantaneous E$_z$ vector distribution at the air/Ag and glass/Ag interfaces of nanogratings (P=340*nm*) at the SRSPPs resonance wavelength of 610*nm*.

**Figure 4. High-resolution and ultra-compact plasmonic subtractive color filters.** (a) A SEM image (i) of plasmonic SCFs with 2, 4, 6, 8 and 10 nanoslits of period P=350*nm*. (ii) and (iii) show the optical microscopy images under TM illumination for the case of 2, 4, 6, 8 and 10 nanoslits with period of 350*nm* and 270*nm*, respectively. (b) Optical microscopy images of cyan (top panel, P=350*nm*) and magenta (bottom panel, P=270*nm*) plasmonic SCFs with 2, 4, 6, 8 and 10 nanoslits of different lengths decreasing from 2*μm* to 0.3*μm*. (c) Top panel shows a SEM image of a plasmonic SCFs mosaic consisting of four different 10×10*μm²* color filter squares (nanogratings with different periods of $P_1$=220*nm*, $P_2$=260*nm*, $P_3$=290*nm*, and $P_4$=350*nm*) with zero separation. Bottom panel is the corresponding optical microscopy image. All of the scale bars are 5*μm*.

**Figure 5. Ultrathin plasmonic subtractive color filters for spectral imaging and transparent displaying.** (a) A SEM image of the fabricated plasmonic subtractive spectroscope with periods gradually changing from 220*nm* to 360*nm* (from left to right, with 1*nm* increment). The line-width of each nanoslit is fixed at 110*nm*. Scale bar, 5*μm*. (b) Optical microscopy image of the plasmonic spectroscope illuminated with TM-polarized white light. (c) Optical microscopy image of (i) a magenta pattern 'L' in a cyan background formed by nanogratings with two different periods ($P_1$=270*nm*, $P_2$=350*nm*) fabricated on a 30*nm*-thick Ag film, illuminated with TM-polarized white light. Scale bar is 10*μm*. (ii) Imaging the background object through the same structure, under TE illumination.

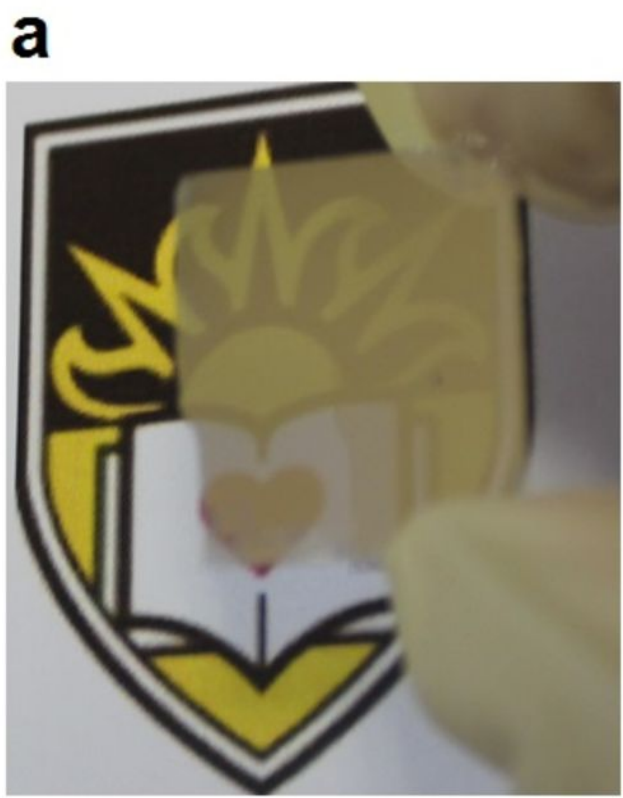
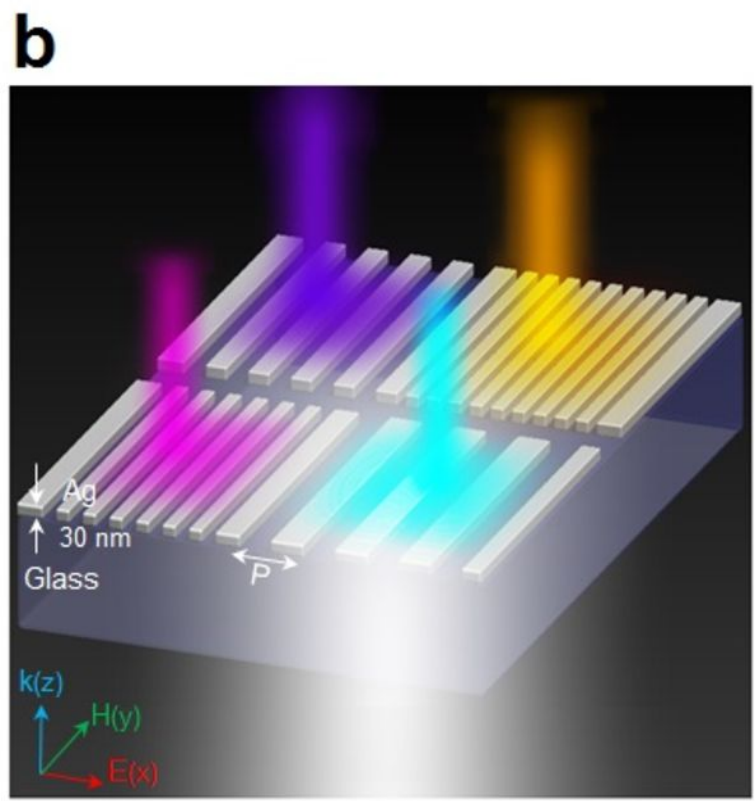
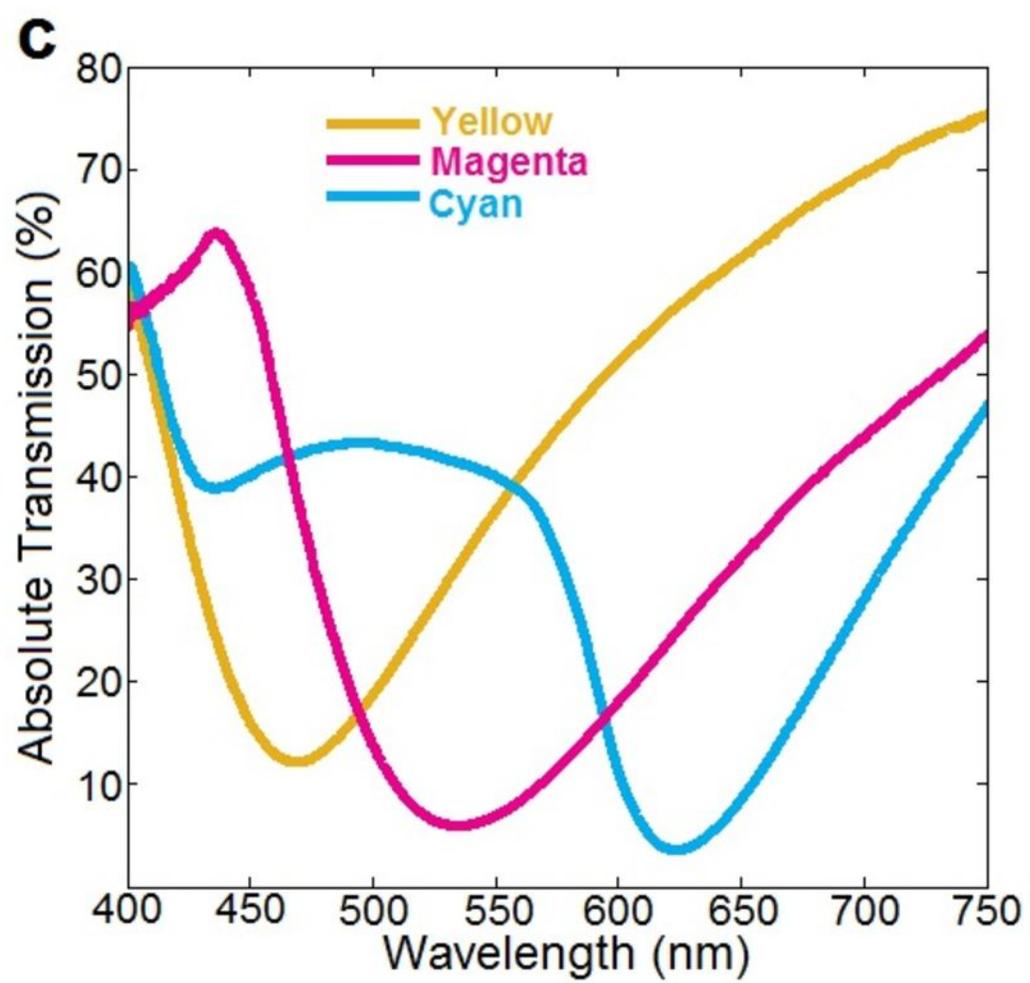
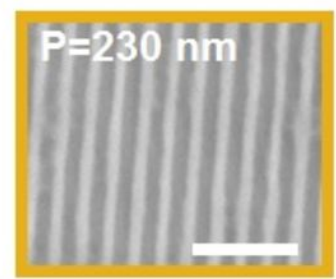
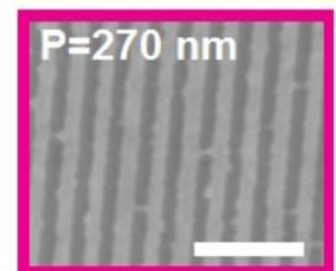
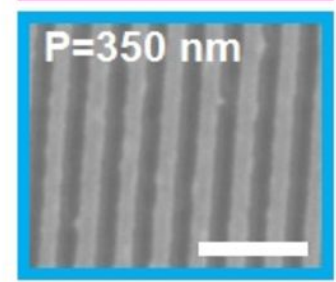

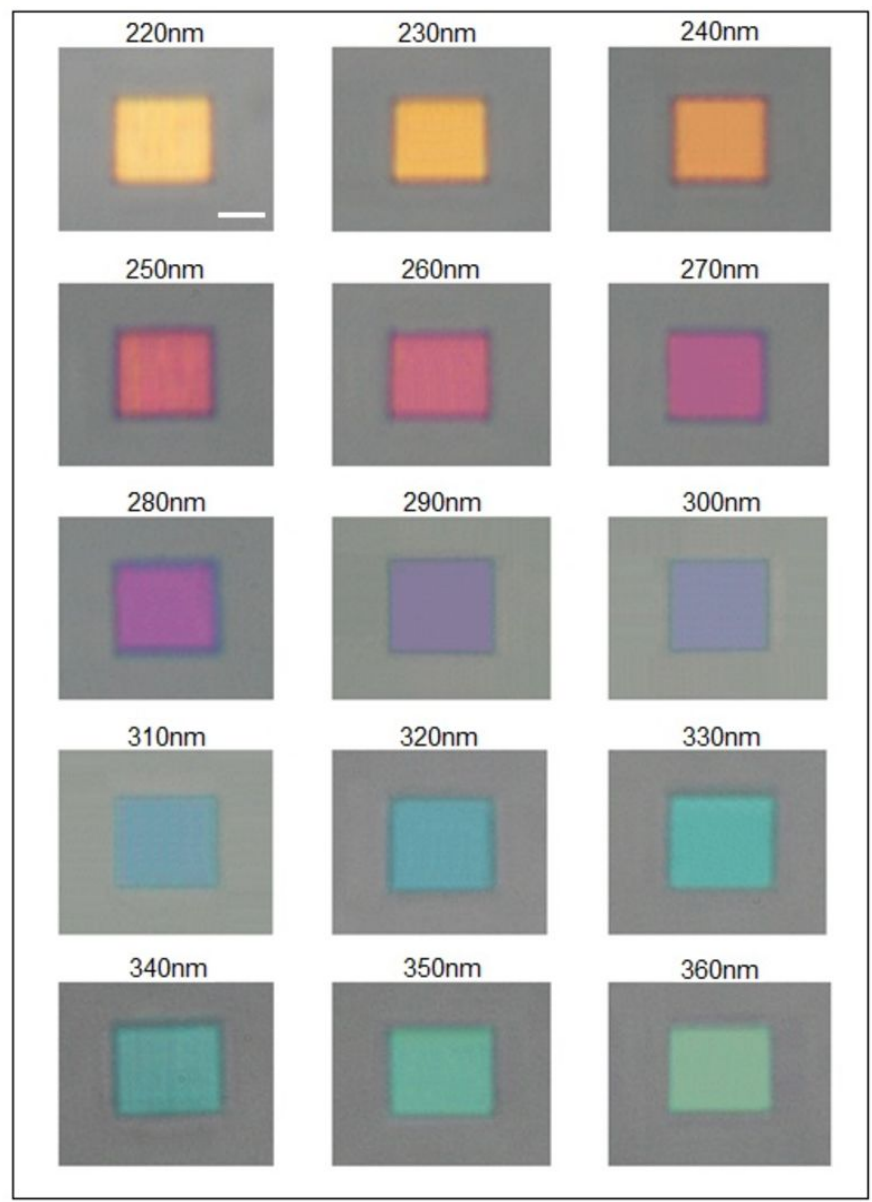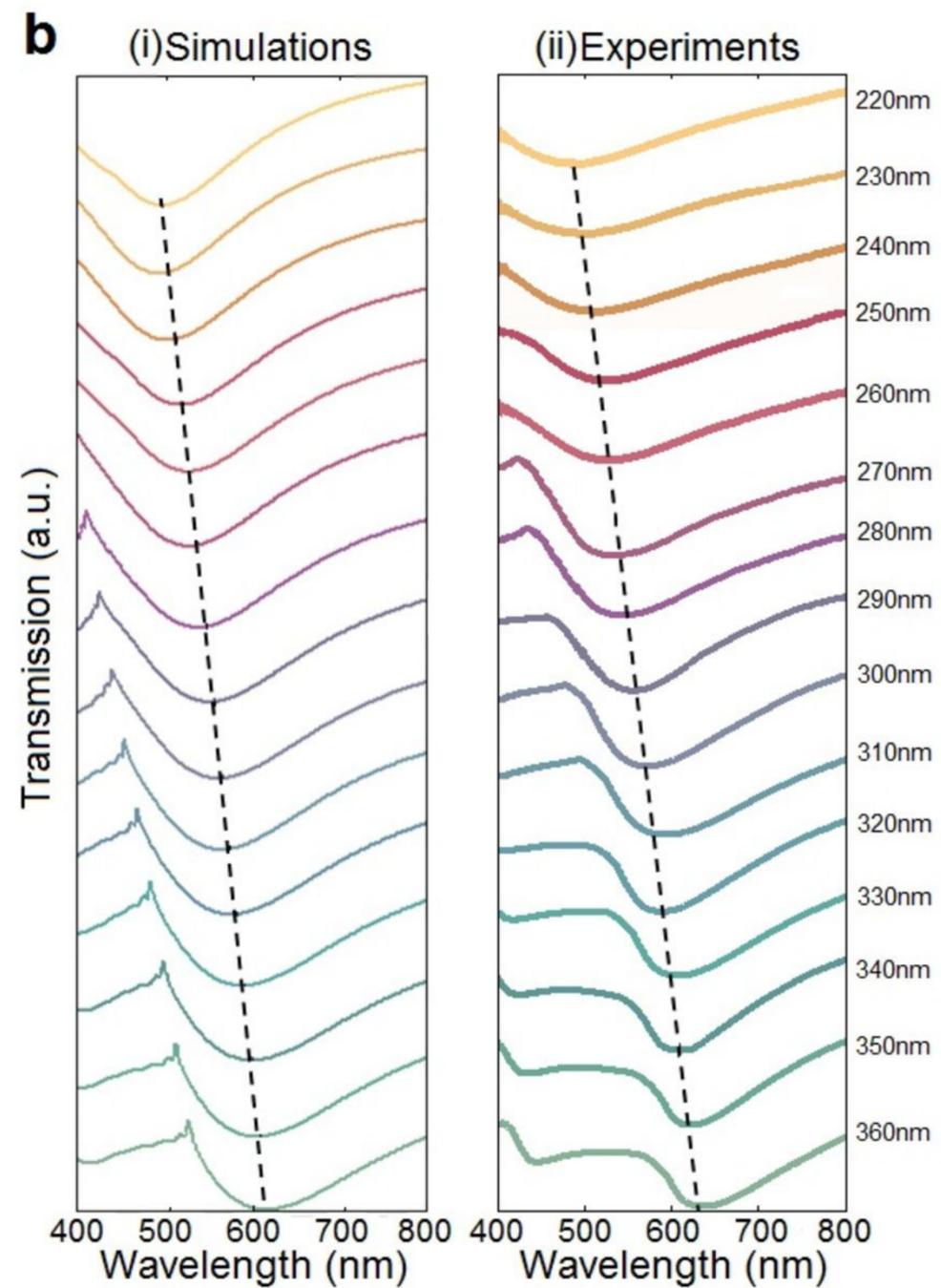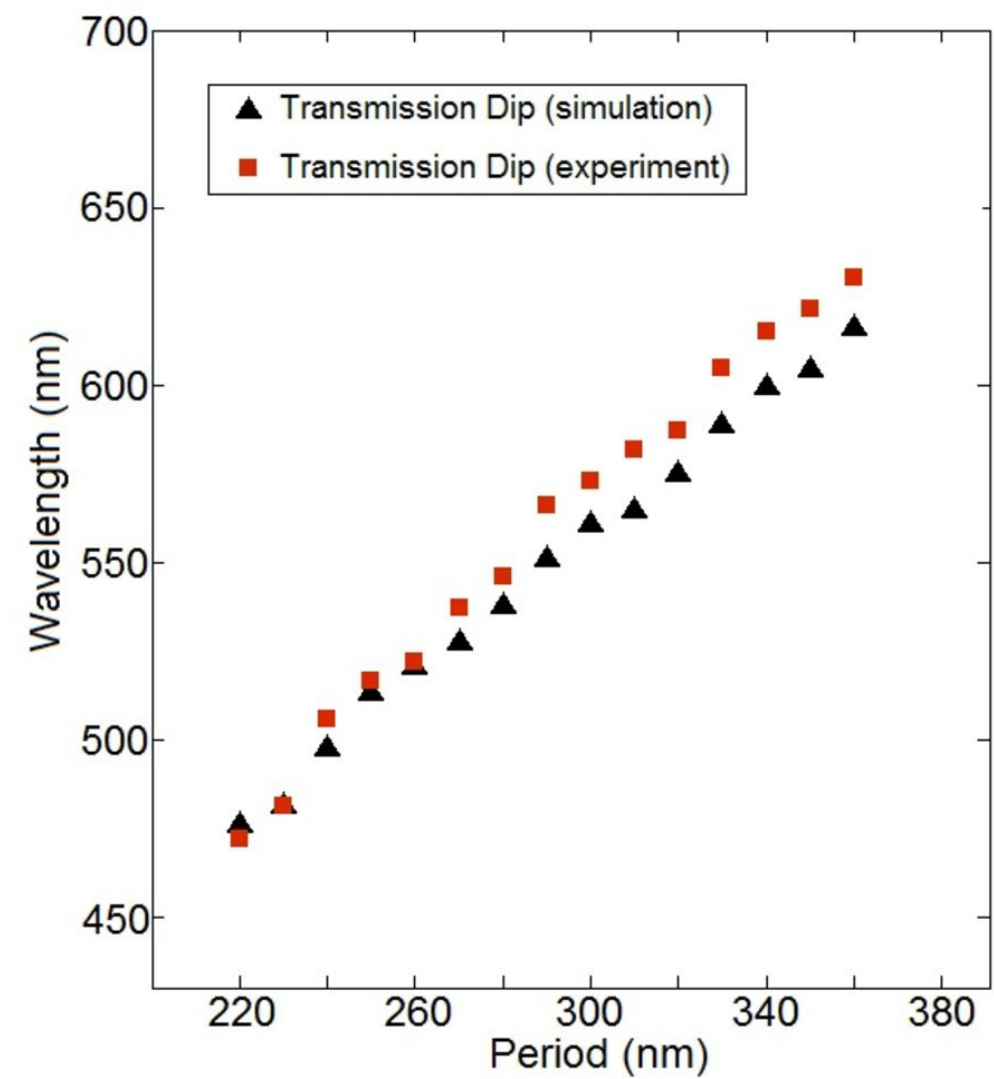

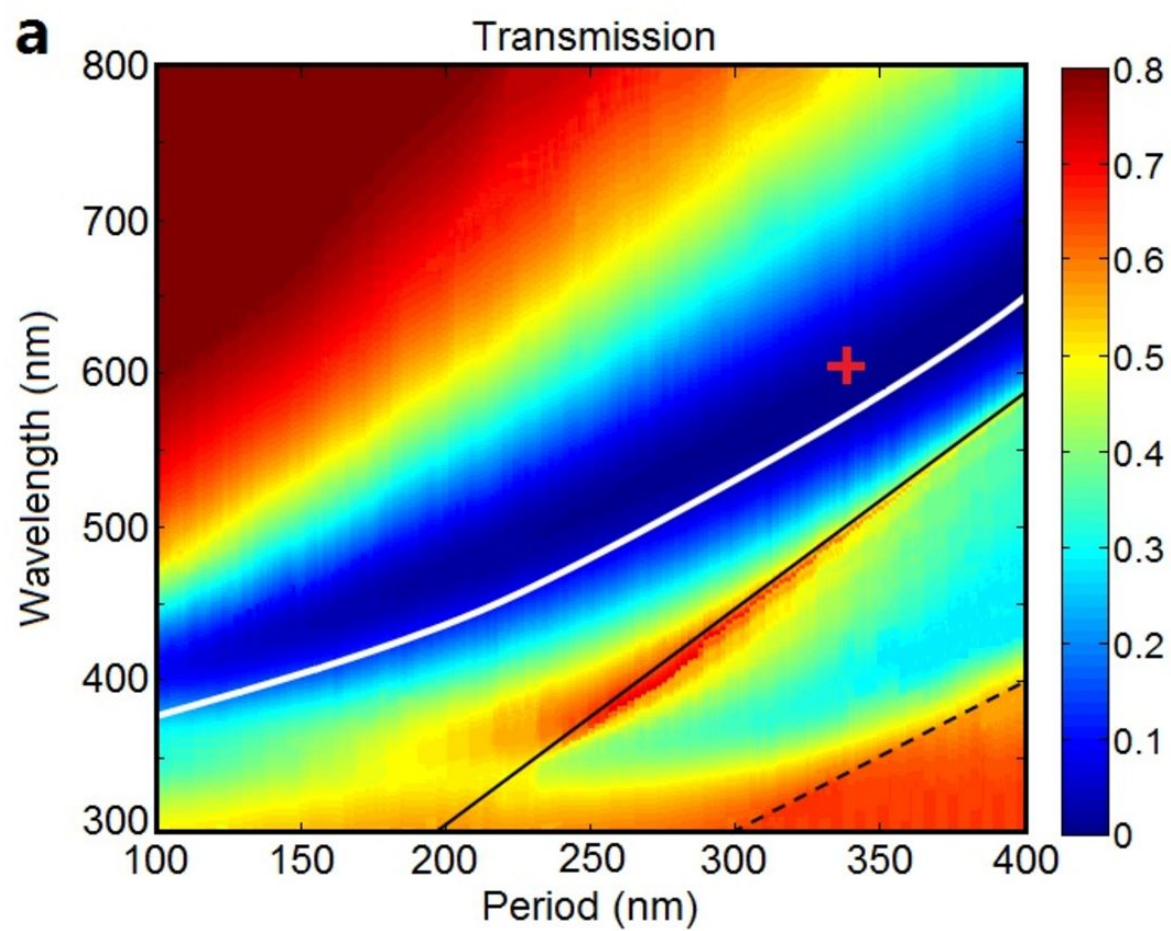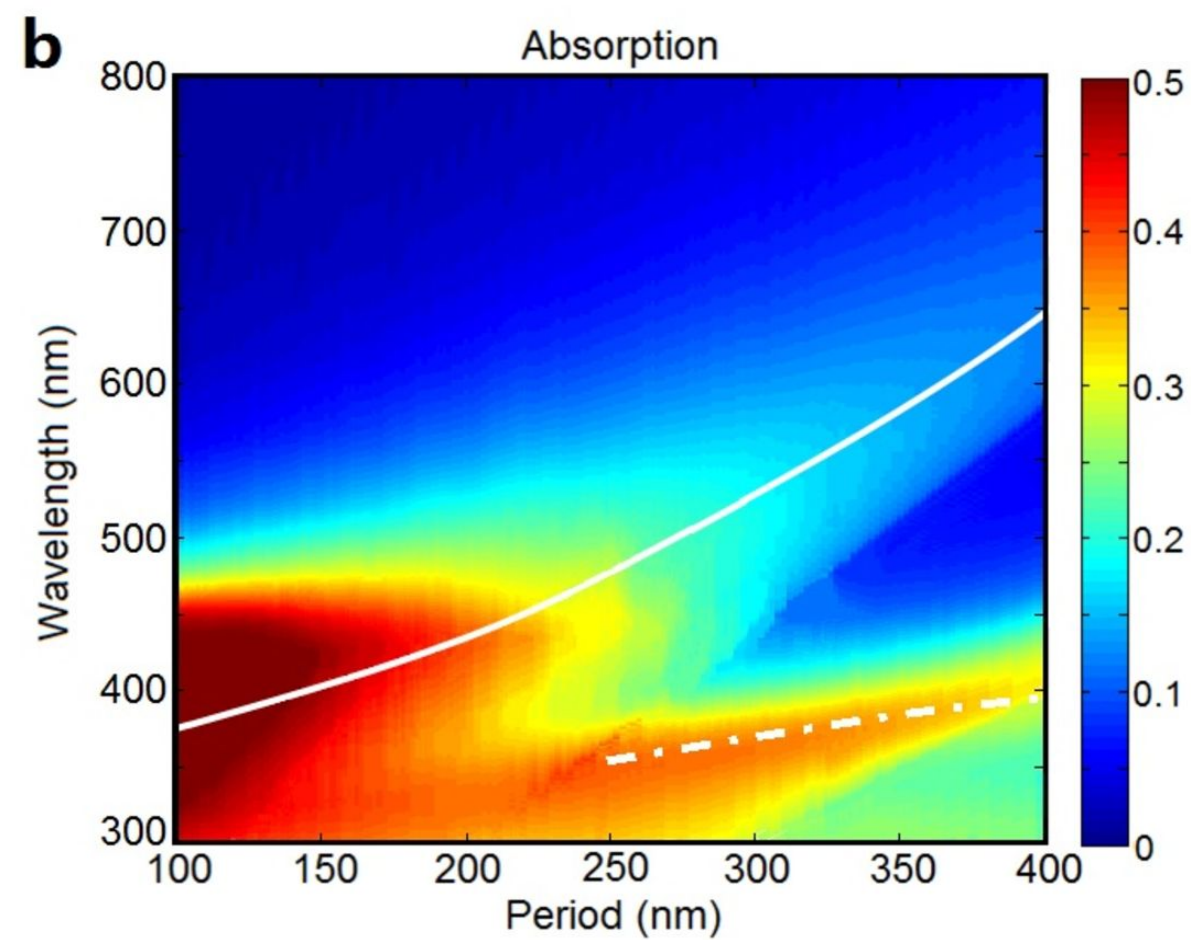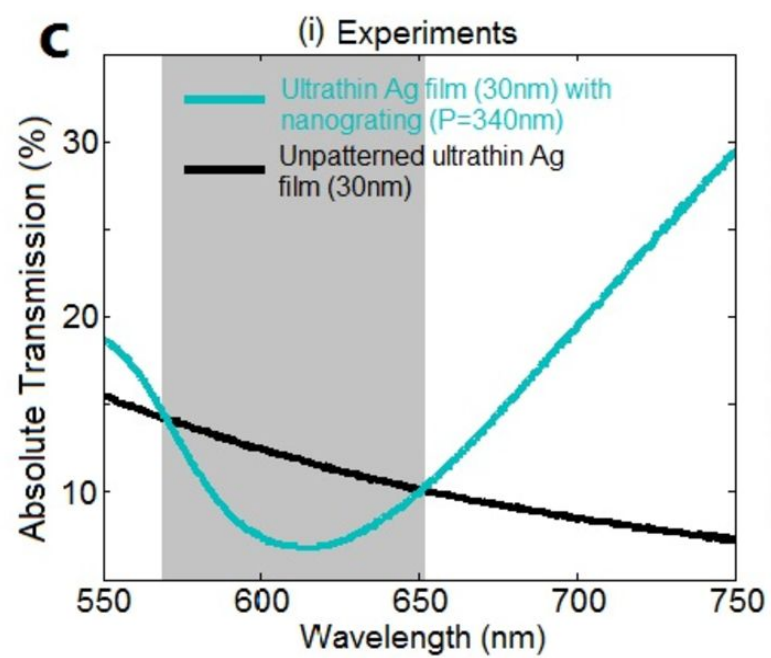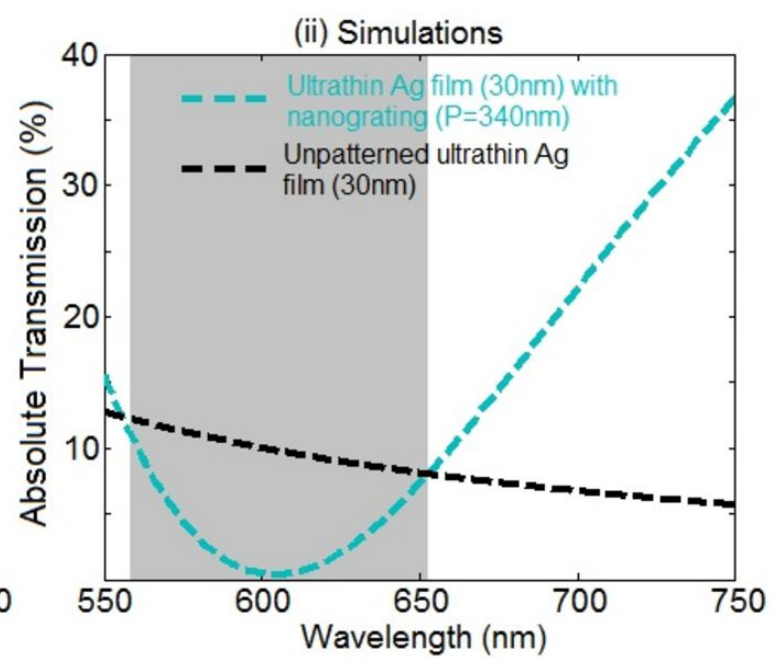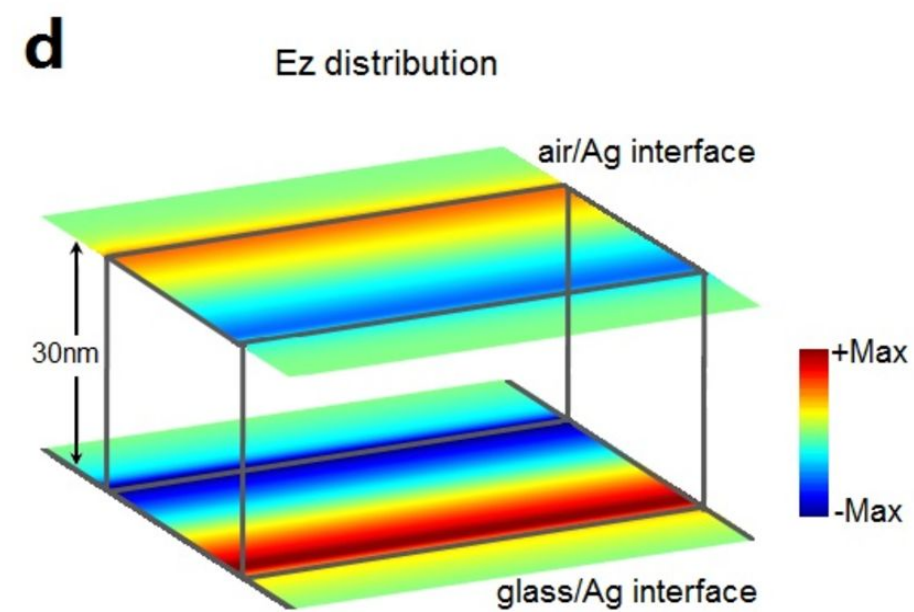

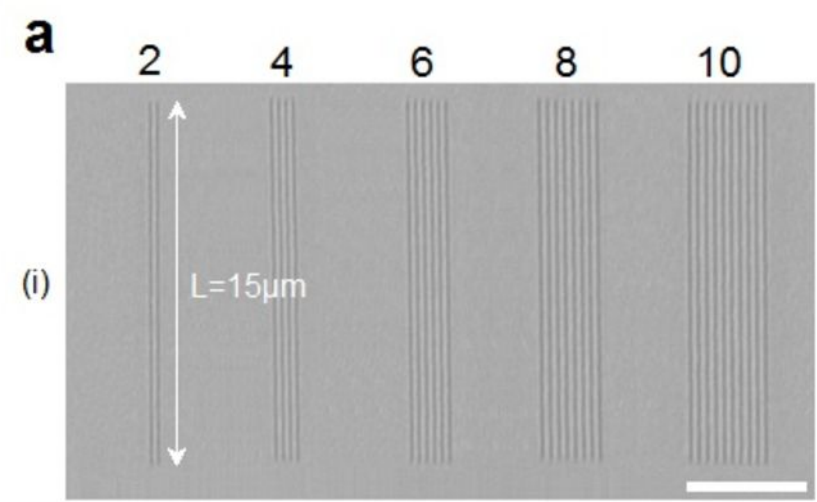
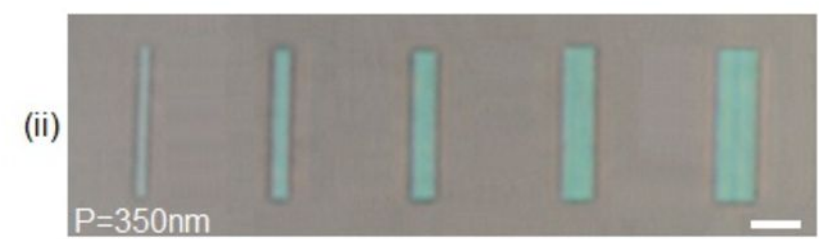
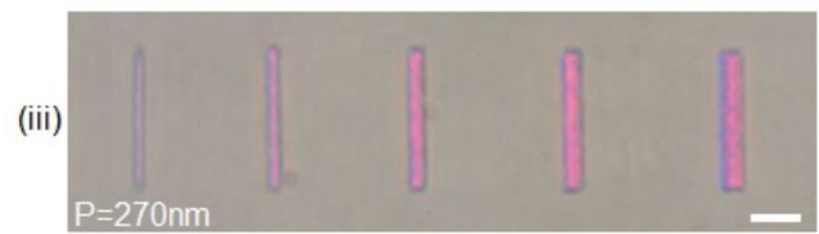
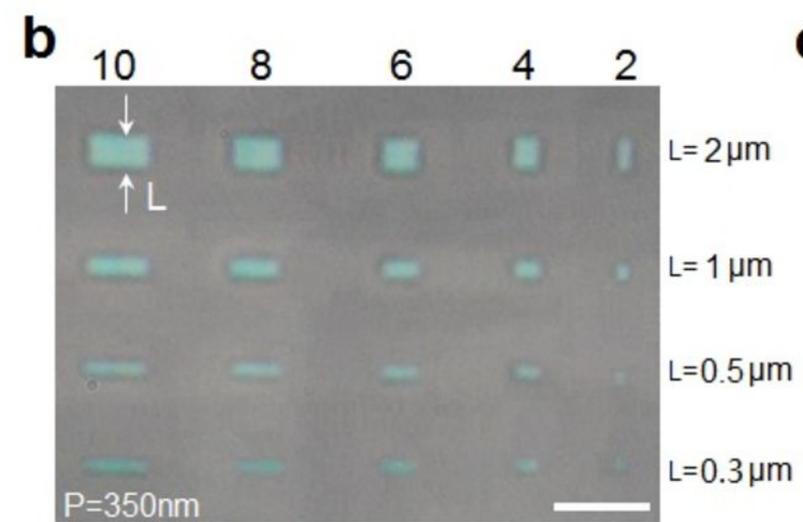
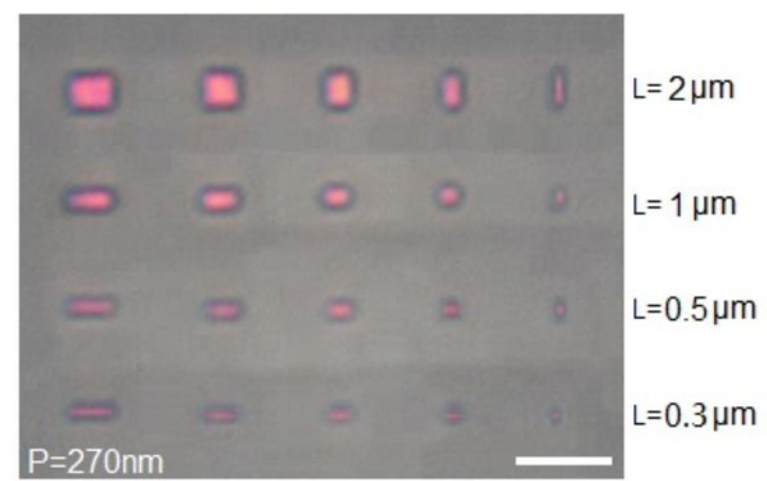
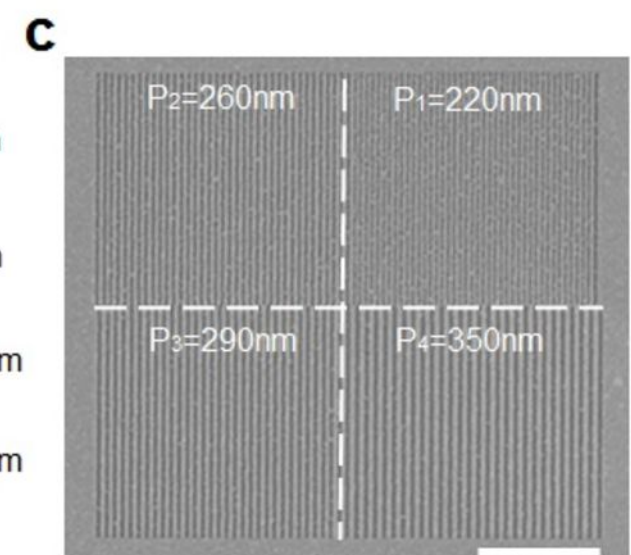
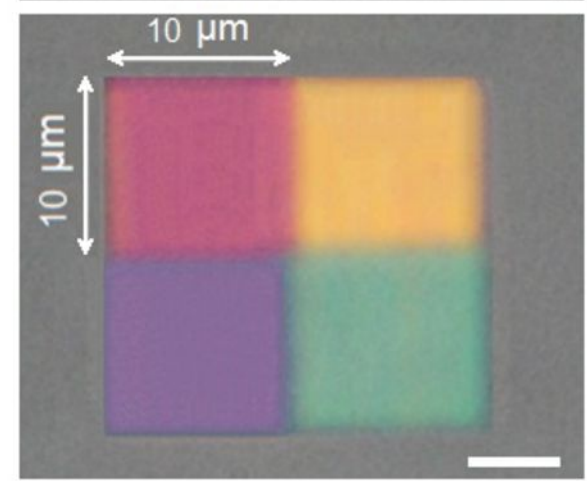

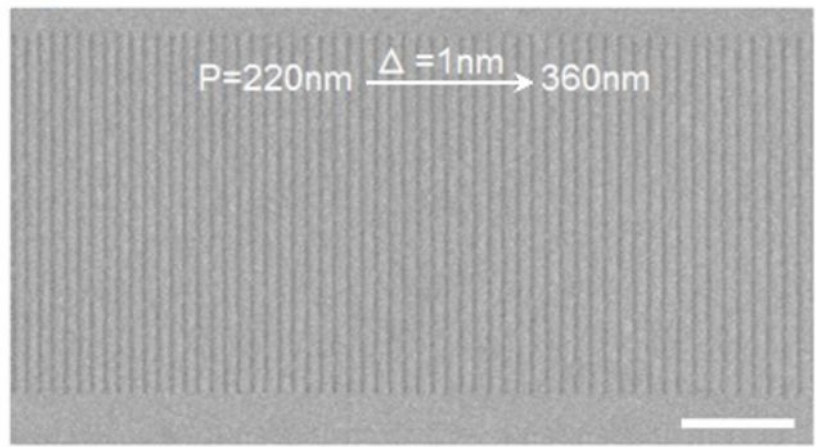
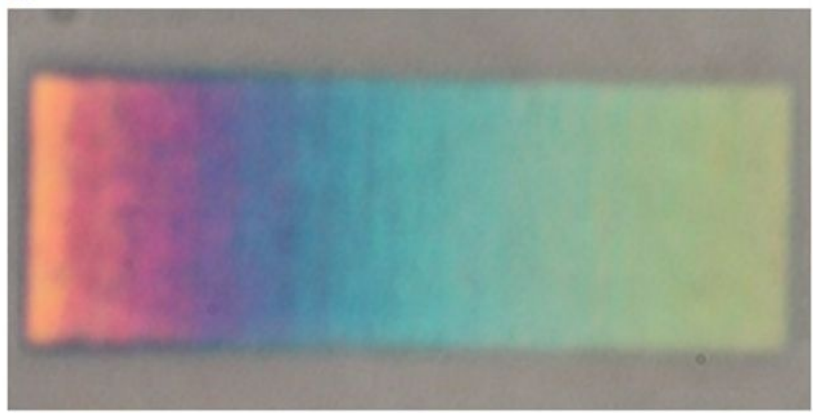
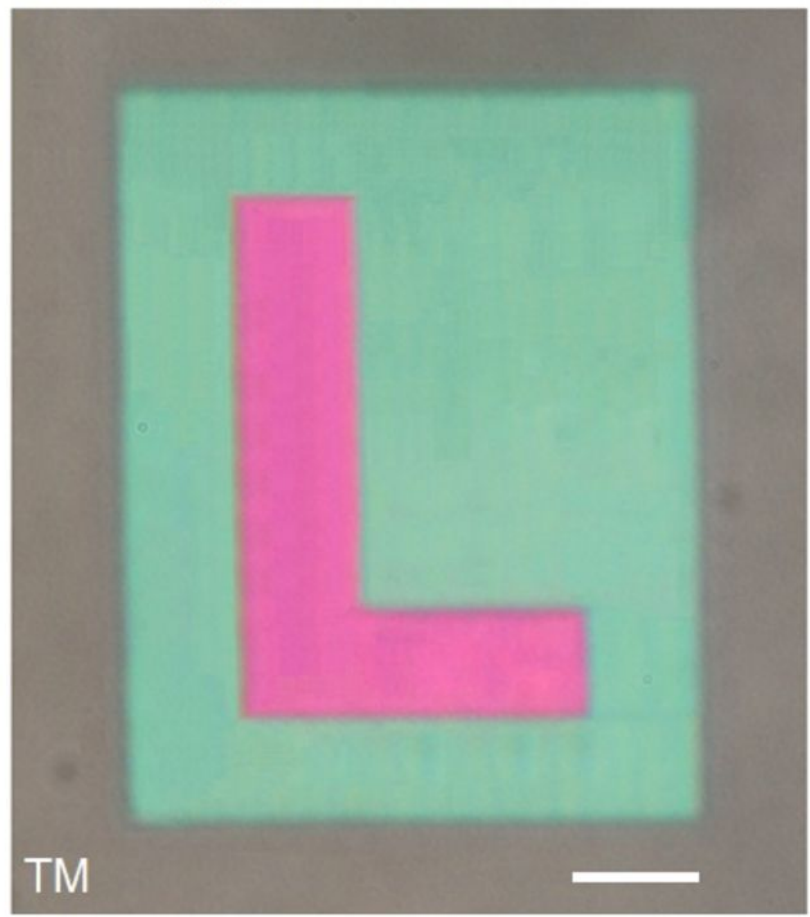
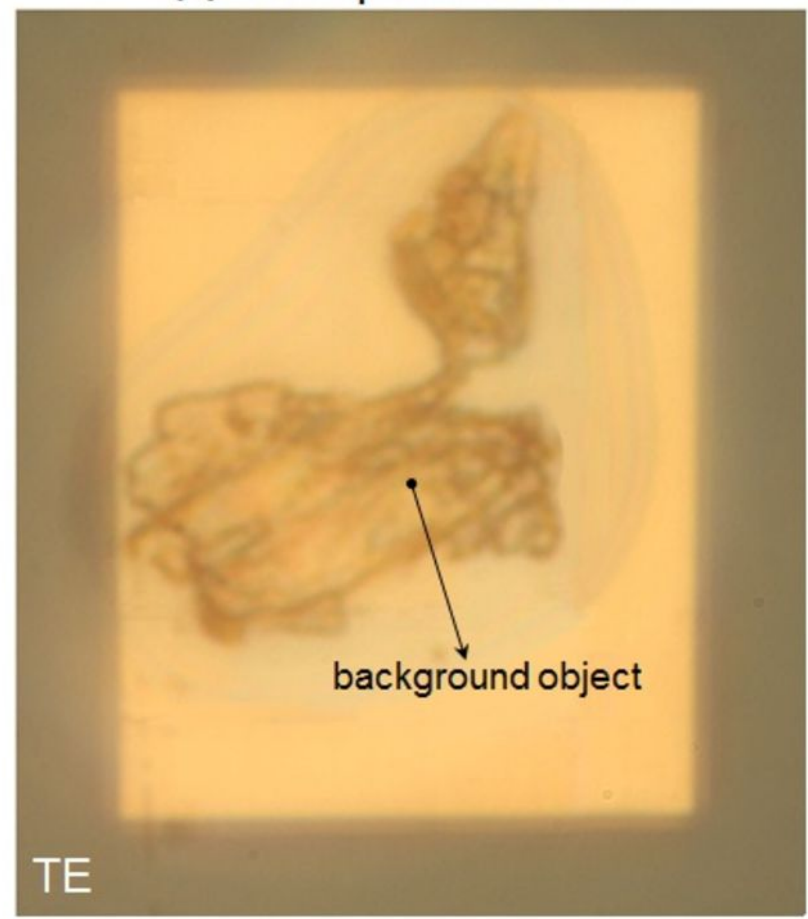